\documentclass[preprint,floatfix] {revtex4}

\usepackage{graphicx}
\usepackage{subfigure}
\usepackage{xcolor}
\usepackage{amsmath}
\usepackage{enumerate}
\usepackage{tabularx}

\usepackage{color, soul}
\definecolor{darkblue}{rgb}{0,0,0.5}
\setulcolor{darkblue}

\begin{document}

\title{A quantum mechanical virial-like theorem for confined quantum systems}

\author{Neetik Mukherjee}
\altaffiliation{Email: neetik.mukherjee@iiserkol.ac.in.}

\author{Amlan K.~Roy}
\altaffiliation{Corresponding author. Email: akroy@iiserkol.ac.in, akroy6k@gmail.com.}
\affiliation{Department of Chemical Sciences\\
Indian Institute of Science Education and Research (IISER) Kolkata, 
Mohanpur-741246, Nadia, WB, India}

\begin{abstract}
Confinement of atoms inside impenetrable (hard) and penetrable (soft) cavity has been studied for nearly eight decades. However, 
a uniform virial theorem for such systems has not yet been found. Here we provide a general virial-like equation in terms of mean 
square and expectation values of potential and kinetic energy operators. It appears to be applicable in \emph{both free and 
confined} situations. Apart from that, a pair of equations has been derived using time independent Schr\"odinger equation, that 
can be treated as a sufficient condition for a given stationary quantum state. Change of boundary condition does not affect these 
virial equations. In \emph{hard} confining condition, the perturbing (confining potential) does not affect the expression; it 
merely shifts the boundary from infinity to a finite region. In \emph{soft} case, on the contrary, the final expression 
includes contributions from perturbing term. These are demonstrated numerically for several representative enclosed systems 
like harmonic oscillator (1D, 3D), hydrogen atom. The applicability in many-electron systems has been discussed. In essence, a 
virial equation has been derived for \emph{free and confined} quantum systems, from simple arguments. 


\vspace{3mm}

\end{abstract}
\maketitle

\section{introduction}
Over the last twenty years confined quantum systems have emerged as a topic of considerable significance for physicists, chemists, 
biologists \cite{sabin2009}. Invention and advancement of contemporary experimental techniques have given the 
required insight about responses of matter under such constrained environments. Furthermore, recent progress in nano-science and 
nano-technology has inspired extensive research activity to explore and acquire more thorough, in-depth understanding. Nowadays, 
various physical, chemical processes are carried out in spatially confined environment. They have profound applications 
in diverse area of research, like condensed matter, semiconductor physics, astrophysics \cite{pang11}, nano-technology, quantum 
dot, wire and well \cite{sen2014electronic}. In recent years, these models are also employed to interpret the trapping of atoms, 
molecules inside fullerene cage, zeolite cavity \cite{sabin2009,sen2014electronic,sen12} etc. 

A quantum particle under the influence of confinement displays many fascinating, distinctive changes in observable physical, 
chemical properties \cite{aquino16,yu17}. Usually, the Schr\"odinger equation (SE) can not be solved 
\emph{exactly}; therefore, one has to take recourse to approximate methods. The perturbative approach leads to an asymptotic 
series \cite{fernandez82}, and standard linear variation method is fraught with the problem of proper boundary behavior, as 
familiar orthonormal basis sets do not vanish at finite boundaries. Thus linear combinations of such bases are explicitly 
inappropriate in representing their eigenstates. Recently for some central potentials (harmonic oscillator, H atom, pseudoharmonic 
oscillator, etc.) under hard confinement condition, such equation can be solved exactly. These eigenfunctions can then be used as 
appropriate orthonormal basis set in other confined systems \cite{ghosal16}.

In 1937, the first model for confined quantum system, a H atom trapped inside an impenetrable barrier was proposed to understand 
its behavior under extreme pressure \cite{michels37}. With time this was found to be somehow restrictive for practical purposes, 
leading to the development of so-called \emph{penetrable} barriers. For sake of convenience, it may be appropriate to categorize 
different confining potentials, following \cite{katriel12}, in two broad classes, namely (i) a \emph{penetrable} potential which 
is finitely bounded from above, whereas in an \emph{impenetrable} case, it rises to infinity at large $r$ (ii) A continuous 
potential will be termed as \emph{smooth} while a \emph{sharp} potential possesses discontinuity. In case of \emph{impenetrable},
\emph{sharp} condition, a potential is modified by the addition of a term that disappear up to a certain distance from origin, 
rising to infinity thereafter. Such potentials are defined as $V=V(r)$ at $0 \leq r \leq r_c$ and $V=\infty$ at $r > r_c$ 
($r_c$ implies confinement radius). In this situation, the Dirichlet boundary condition $R_{n,\ell} (0)=R_{n,\ell} (r_c) =0$ is 
obeyed \cite{sabin2009}. On the other hand, an \emph{impenetrable, smooth} potential is defined as $V=V(r)+V_c(r)$, where 
$V_c(r)$ is the confining potential that becomes infinity at $r \rightarrow \infty$, and remain continuous otherwise 
\cite{wilson94,patil09}. Similarly, for \emph{penetrable, sharp} case the potential has the form: $V=V(r)$ at $0 \leq r \leq r_c$ 
and $V=V_c (r) $ at $r > r_c$, where, $V_c(r)$ is the confining potential \cite{koo79}. Finally, in the \emph{penetrable, smooth}
case it becomes, $V=V(r)+V_c(r)$ \cite{adamowski00}. In recent years, various models were proposed and investigated by many 
authors \cite{sen2014electronic,aquino13,sen05,randazzo16, cabrera16}, especially in the context of H atom, maintaining these 
confinement conditions, revealing numerous striking features \cite{sabin2009,sen2014electronic,stevanocic10,montgomery12,
cabrera13}.                

Extensive theoretical calculations have been made in case of confined harmonic oscillator (CHO) (1D, 2D, 3D, $d$ dimension) 
\cite{coll17,aquino97, campoy2002,montgomery07,roy14, ghosal16} and confined hydrogen atom (CHA) inside an impenetrable 
cavity \cite{coll17,goldman92,aquino95,garza98,laughlin02,burrows06, aquino07cha,baye08,ciftci09,sen2014electronic,roy15a,
centeno17}. They offer many extraordinary features, especially relating to \emph{simultaneous, incidental, inter-dimensional 
degeneracy} \cite{montgomery07} in their energy spectra. Effect of contraction on ground and excited energy states, as well as 
other properties like hyperfine splitting constant, dipole shielding factor, nuclear magnetic screening constant, static and 
dynamic polarizability, etc., were explored \cite{sabin2009,sen2014electronic, sen12}. A wide range of attractive analytical and 
numerical approaches including perturbation theory, Pad\'e approximation, WKB method, Hypervirial theorem, power-series solution, 
supersymmetric quantum mechanics, Lie algebra, Lagrange-mesh, asymptotic iteration, generalized pseudo-spectral method, etc., 
were invoked to solve the relevant eigenvalue problem \cite{goldman92,aquino95,garza98,laughlin02,burrows06,aquino07cha,baye08,
ciftci09,roy15a}. Exact solutions \cite{burrows06} of CHA are expressible in terms of Kummer M-function (confluent hypergeometric).    

In quantum mechanics, stationary states of a bound system satisfy the virial theorem (VT). In fact, it is a necessary condition 
for a quantum stationary state to follow \cite{lowdin59}. Historically the quantum mechanical VT was derived from analogy with
classical counterpart; for a non-relativistic Hamiltonian, it offers a relation between expectation values of kinetic energy 
and directional derivatives of potential energy. {\color{red}In this regard, it is important to point out that a variationally optimized 
wave function also obeys the VT. Hence, it becomes a necessary condition for an exact wave function to obey. On the contrary, 
obeying this relation will not ensure that the state to be exact.} After some controversy, it is now generally accepted that the standard 
form of VT is not obeyed in enclosed condition; rather a modified form is invoked. Several attempts were made to find an 
appropriate form of VT in such systems \cite{cotrell51,brown58,fernandez82}. Previously, some semi-classical strategies based on 
Wilson-Sommerfeld rule and uncertainty principle were also adopted to construct VT in such systems \cite{mukhopadhyay05}. In 
recent years, standard form of VT and Hellmann-Feynman theorem (HFT) were combined to design new virial-like expression for 
\emph{penetrable and impenetrable} CHA \cite{katriel12}, however, the mathematical forms of the expressions change from system 
to system. {\color{red}Importantly, all these relations can only serve as necessary condition for an exact state to obey.} In this endeavour our aim 
is to design a uniform virial-like expression for both \emph{free and confined} conditions using time-independent SE, {\color{red}the Hyper-virial 
theorem (HVT) \cite{hirsh60},} along with mean square values and expectation values of potential and kinetic 
energy operators. {\color{red}Apart from that, a new relation involving SE and HVT has been derived, which can serve as a sufficient condition (only true for exact states) 
for a bound stationary state to obey.} In this scenario, detailed derivation of these relations are given in 
Sec.~II. Next we proceed to verify the utility and applicability of these relations in the context of several representative 
confined systems. We begin Sec.~III with the oldest, primitive model of \emph{hard} confinement, where the potential is trapped 
inside an infinite wall satisfying the \emph{Dirichlet boundary condition}. In this category, at first, we discuss the typical 
and most prolific cases of CHO (1D, 3D) as well as a CHA. Later, this is extended to the so-called shell-confined H-atom (SCHA), 
in order to understand the role of nodal structure in confined condition. This can be potentially treated as a confined 
off-centre model, needed to probe quantum wells/dots. With time, a new model for off-centre quantum dot structures was also 
adopted, but within the frame work of \emph{Newmann boundary condition}, a prominent examples being the trapping of H atom inside 
a \emph{homogeneous, impenetrable} cavity (HICHA). It my be noted that, at $r_c \rightarrow 0$ this behaves similar to CHA, while 
at $r_c \rightarrow \infty$ it resembles a free H-atom (FHA). In order to make these artificial atomic models more realistic, a 
finite wall was placed at a certain $r_c$; this has been widely used to study the properties of encapsulated atoms within 
fullerene cage and zeolite cavity. As an approximation to this, we explore the case of a H atom inside an \emph{inhomogeneous, 
penetrable} spherical cavity (SPCHA). Apart from that, to incorporate the interaction of particle with the environment 
\emph{homogeneous, penetrable} confinement model was proposed--for this we consider an H atom under similar condition (HPCHA). 
This will help us about the advantages of presently derived relations in the pursuit of confined quantum systems. Section IV 
makes a few concluding remarks. 
 
\section{Theoretical Formalism}
The time-independent non-relativistic SE for a system may simply be written as,
\begin{equation}
(\hat{T}+\hat{V}) \psi_{n}(\tau)= \mathcal{E}_{n} \psi_{n} (\tau), 
\end{equation}
where, $\hat{T}, \hat{V}$ are usual kinetic and potential energy operators, while $\tau$ is a generalized variable. 
After some straightforward algebra (multiplying both sides by $\hat{T}$, integrating over whole space and rearranging), one gets, 
\begin{equation}
\langle \hat{T}^{2}\rangle_{n} + \langle \hat{T} \hat{V} \rangle_{n} = \mathcal{E}_{n} \langle \hat{T} \rangle_{n}. 
\end{equation}
Now, replacing $\mathcal{E}_{n}=\langle \hat{T} \rangle_{n} + \langle \hat{V} \rangle_{n}$ in Eq.~(2) produces, 
\begin{equation}
\langle \hat{T}^{2} \rangle_{n}-\langle \hat{T} \rangle^{2}_{n}	= \langle \hat{V} \rangle_{n} \langle \hat{T} \rangle_{n}
-\langle \hat{T}\hat{V} \rangle_{n}. 
\end{equation}
A similar consideration using $\hat{V}$ leads to the following equation, 
\begin{equation}
\langle \hat{V}^{2} \rangle_{n}-\langle \hat{V} \rangle^{2}_{n} = \langle \hat{T} \rangle_{n} \langle \hat{V} \rangle_{n}
-\langle \hat{V}\hat{T} \rangle_{n}
\end{equation}	
From hypervirial theorem, it can be proved that, $\langle \hat{T}\hat{V} \rangle_{n}=\langle \hat{V}\hat{T} \rangle_{n}$. 
Hence, from Eqs.~(3)-(4), one obtains, 
\begin{equation}
\begin{aligned}
\langle \hat{T}^{2} \rangle_{n}-\langle \hat{T} \rangle^{2}_{n} & = \langle \hat{V}^{2} \rangle_{n}-\langle \hat{V} 
\rangle^{2}_{n} \\
(\Delta \hat{T}_{n})^{2} = (\Delta \hat{V}_{n})^{2} & = \langle \hat{V} \rangle_{n} \langle \hat{T} \rangle_{n}
-\langle \hat{T}\hat{V} \rangle_{n} = \langle \hat{T} \rangle_{n} \langle \hat{V} \rangle_{n}
-\langle \hat{V}\hat{T} \rangle_{n}.
\end{aligned}
\end{equation}
This relation suggests that, the magnitude of error incurred in $\langle \hat{T} \rangle_{n}$ and $\langle \hat{V} \rangle_{n}$
are equal. Now, one can easily interpret the fact that, $\mathcal{E}_{n}$ is a sum of two average quantities but still provides 
exact result. It is due to the cancellation of errors between $\langle \hat{T} \rangle_{n}$ and $\langle \hat{V} \rangle_{n}$. 

Interestingly, using the condition $\langle \hat{T}\hat{V} \rangle_{n}=\langle \hat{V}\hat{T} \rangle_{n}$ and exploiting Eqs.~
(3) and (4), one can reach the expression, 
\begin{equation}
\langle \hat{T}^{2} \rangle_{n} = \mathcal{E}_{n}\left(\mathcal{E}_{n}-2\langle \hat{V} \rangle_{n} \right) 
+ \langle \hat{V}^{2} \rangle_{n}. 
\end{equation}
Thus, instead of performing the fourth order derivative of $\psi_{n} (\tau)$, one can alternatively evaluate 
$\langle \hat{T}^{2} \rangle_{n}$ from a knowledge of $\mathcal{E}_{n}$, $\langle \hat{V} \rangle_{n}$ and 
$\langle \hat{V}^{2} \rangle_{n}$. 

Now we wish to verify whether Eq.~(5) is true for eigenstates only or not. Let us consider two functions having forms  
$ \phi_{1}=|\hat{T}-\langle \hat{T} \rangle_{n}| \psi_{n} \rangle $ and $ \phi_{2}=|\hat{V}-\langle \hat{V} \rangle_{n}| \psi_{n} 
\rangle$. Making use of Schwartz inequality, it is possible to write, 
\begin{equation}
\begin{aligned}
\langle \phi_{1}|\phi_{1} \rangle \langle \phi_{2}| \phi_{2} \rangle & \geq |\langle \phi_{2}| \phi_{1} \rangle|^{2} \\	
(\Delta \hat{T}_{n})^{2} (\Delta \hat{V}_{n})^{2}  & \geq |\langle \hat{T} \hat{V} \rangle_{n} - 
\langle \hat{T} \rangle_{n} \langle \hat{V} \rangle_{n}|^{2}
\end{aligned}
\end{equation}
This inequality becomes equality when $\phi_{1}$ and $\phi_{2}$ are linearly dependent. That implies,  
\begin{equation}
|\hat{T}-\langle \hat{T} \rangle_{n}| \psi_{n} \rangle = \alpha |\hat{V}-\langle \hat{V} \rangle_{n}| \psi_{n} \rangle, 
\end{equation}
where $\alpha$ is a number. Putting this back in the inequality and doing some algebraic rearrangement, we get,   
\begin{equation}
\begin{aligned}
\alpha^{2} (\Delta \hat{T}_{n})^{2} (\Delta \hat{V}_{n})^{2}  & = |\langle \hat{T} \hat{V} \rangle_{n} - 
\langle \hat{T} \rangle_{n} \langle \hat{V} \rangle_{n}|^{2}  \\
\alpha^{2} & =\frac{|\langle \hat{T} \hat{V} \rangle_{n} - \langle \hat{T} \rangle_{n} \langle \hat{V} \rangle_{n} |^{2}}
{(\Delta \hat{T}_{n})^{2} (\Delta \hat{V}_{n})^{2}}. 
\end{aligned}
\end{equation}
Choice of $\alpha^{2}=1$ yields the following expression,
\begin{equation}
(\Delta \hat{T}_{n} ^{2}) (\Delta \hat{V}_{n} )^{2} = |\langle \hat{T} \hat{V} \rangle_{n} - 
\langle \hat{T} \rangle_{n} \langle \hat{V} \rangle_{n}|^{2} 
\end{equation}
Here $\alpha^{2}=1$. Now, left multiplying Eq.~(8) by $\langle \psi_n|(T- \langle T \rangle_n)|$, followed by integration
over whole space and rearrangement leads to,  
\begin{equation}
(\Delta \hat{T}_{n} )^{2} = (\Delta \hat{V}_{n} )^{2} = |\langle \hat{T} \hat{V} \rangle_{n} - 
\langle \hat{T} \rangle_{n} \langle \hat{V} \rangle_{n}|
\end{equation}
Equation~(11) is valid for two values of $\alpha$, namely, 1 or $-1$. When $\alpha=-1$,
\begin{equation}
\begin{aligned}
|\hat{T}-\langle \hat{T} \rangle_{n}| \psi_{n} \rangle & = - |\hat{V}-\langle \hat{V} \rangle_{n}| \psi_{n} \rangle  \\
(\hat{T}+\hat{V})|\psi_{n}\rangle & = \left(\langle \hat{T} \rangle_{n} + \langle \hat{V} \rangle_{n} \right) |\psi_{n}\rangle 
\end{aligned}
\end{equation}
Which is nothing but Schr\"odinger Equation: $ \hat{H} |\psi_{n} \rangle  = \mathcal{E}_{n}|\psi_{n} \rangle $. 
Whereas $\alpha=1$ gives, 
\begin{equation}
(\hat{T}-\hat{V})|\psi_{n}\rangle = [\langle \hat{T} \rangle_{n}-\langle \hat{T} \rangle_{n}]|\psi_{n} \rangle, 
\end{equation}
which does not concern us here. 

This above discussion suggests that, Eq.~(11) is a necessary condition for a stationary state to obey and Eq.~(5) is a special 
case of it. Now, to verify the suitability of Eq.~(5), it is useful to study $\langle \hat{H}^{2} \rangle_{n}- 
\langle \hat{H} \rangle^{2}_{n}= \left(\Delta \hat{H}_{n}\right)^{2}$, because, only for eigenstates it is \emph{zero}. Thus,   
\begin{equation}
(\Delta \hat{H}_{n})^{2}= (\Delta \hat{T}_{n} )^{2} + (\Delta \hat{V}_{n} )^{2} +
[\langle \hat{T} \hat{V} \rangle_{n} - \langle \hat{T} \rangle_{n} \langle \hat{V} \rangle_{n}] +
[\langle \hat{V} \hat{T} \rangle_{n} - \langle \hat{T} \rangle_{n} \langle \hat{V} \rangle_{n}]
\end{equation} 
Now, putting the condition of Eq.~(5) in Eq.~(14) one can obtain,
\begin{equation}
\left(\Delta \hat{H}_{n}\right)^{2}=0
\end{equation}
This clearly states that, Eq.~(5) is a \emph{sufficient} condition for an eigenfunction to obey. Hence, once this relation is 
obeyed by $\psi_{n}$, it is an eigenfunction of that particular $\hat{H}$. But $(\Delta \hat{T}_{n})^{2}=(\Delta 
\hat{V}_{n})^{2}$ is a necessary condition for a quantum system to obey, which is actually a virial-like expression. Now, it 
will be interesting to examine the applicability of Eq.~(5) in the context of confined quantum systems.  

For our current purpose at hand, without loss of generality, our relevant radial SE under the influence of confinement is, 
\begin{equation}
\left[-\frac{1}{2} \ \frac{d^2}{dr^2} + \frac{\ell (\ell+1)} {2r^2} + v(r) +v_c (r) \right] \psi_{n,\ell}(r)=
\mathcal{E}_{n,\ell} \ \psi_{n,\ell}(r),
\end{equation}
where $v(r)$ signifies the \emph{unperturbed effective} potential (for example, in a many-electron system that may include 
effective electron-nuclear attraction and electron-electron repulsion), and our desired confinement inside a spherical cage 
is accomplished by invoking the potential $v_c(r)$, with $\hat{V}=v(r)+v_{c}(r)$. Thus in a confinement scenario, validity of 
Eq.~(5) can be checked by deriving the expressions of $\langle \hat{T} \hat{V} \rangle_{n,\ell}$, $\langle \hat{V} 
\hat{T} \rangle_{n,\ell}$, $\langle \hat{V}^{2} \rangle_{n,\ell}$ and $\langle \hat{V} \rangle_{n,\ell}$
(other integrals remain unchanged). Towards this end, Eq.~(5) may be modified as follows: 
\begin{equation}
(\Delta \hat{T}_{n,\ell} )^{2}  =  \langle \hat{T}^{2} \rangle_{n,\ell}-\langle \hat{T} \rangle^{2}_{n,\ell},
\end{equation}
\begin{equation}
\begin{aligned}
(\Delta \hat{V}_{n,\ell})^{2} = \langle v(r)^{2} \rangle_{n,\ell} + \langle v(r)v_{c}(r) \rangle_{n,\ell} + 
\langle v_{c}(r)v(r) \rangle_{n,\ell} + \langle v_{c}(r)^{2} \rangle_{n,\ell} -  \langle v(r) \rangle^{2}_{n,\ell} - 
\langle v_{c}(r) \rangle^{2}_{n,\ell} \\
-2\langle v(r) \rangle_{n,\ell} \langle v_{c}(r) \rangle_{n,\ell},
\end{aligned}
\end{equation}
\begin{equation}
\begin{aligned}
\langle \hat{T} \rangle_{n,\ell} \langle \hat{V} \rangle_{n,\ell}-\langle \hat{V}\hat{T} \rangle_{n,\ell} = 
\langle \hat{T} \rangle_{n,\ell} \left[ \langle v(r) \rangle_{n,\ell}+\langle v_{c}(r) \rangle_{n,\ell} \right] - 
\langle \hat{T} v(r) \rangle_{n,\ell} - \langle \hat{T}v_{c}(r) \rangle_{n,\ell}  \\
\langle \hat{T} \rangle_{n,\ell} \langle \hat{V} \rangle_{n,\ell}-\langle \hat{T}\hat{V} \rangle_{n,\ell} = 
\langle \hat{T} \rangle_{n,\ell} 
\left[ \langle v(r) \rangle_{n,\ell}+\langle v_{c}(r) \rangle_{n,\ell} \right] - \langle v(r)\hat{T} \rangle_{n,\ell} - 
\langle v_{c}(r)\hat{T} \rangle_{n,\ell}. 
\end{aligned}
\end{equation}
In what follows, we will analyze the above-mentioned criteria for a number of important confining potentials, as mentioned in 
the introduction section, \emph{viz.}, (i) CHO in 1D and 3D (ii) A H atom encapsulated in five different confining 
environments, namely, CHA, SCHA, HICHA, SPCHA and HPCHA. This will offer us the opportunity to understand the effect of 
boundary condition on derived relations. It may be recalled that, out of these seven different potentials, 1DCHO, 3DCHO 
and CHA are exactly solvable. However, it is instructive to note that, in order to construct the exact wave function for a 
specific state, one needs to supply energy eigenvalue, which is calculated using imaginary-time propagation \cite{roy15} and 
generalized pseudo-spectral \cite{roy04, sen06, roy13, roy14, roy15a} method respectively for 1D and 3D problems. Except CHA, 
in all the remaining confining H atom cases, we have employed numerically calculated wave functions and energies through GPS
scheme. Now, we can use relation in Eq.~(5) to inspect the goodness of numerical wave function.

\begingroup           
\squeezetable
\begin{table}
\caption{$\mathcal{E}_{n}, \left(\Delta V_{n}\right)^{2}, \left(\Delta T_{n}\right)^{2}, \langle T \rangle_{n}\langle V 
\rangle_{n}$ $-$ $\langle TV \rangle_{n}, \langle T \rangle_{n}\langle V \rangle_{n}$ $-$ $\langle VT \rangle_{n}$ values for 
$n=0,1$ states in 1D CHO at six selected values of $x_{c}$, namely $0.1, 0.5, 1, 3, 5, \infty$. See text for detail.}
\centering
\begin{ruledtabular}
\begin{tabular}{l|l|llllll}
$n$ & Property           &  $x_c=0.1$ & $x_c=0.5$ & $x_c=1$ & $x_c=3$ & $x_c=5$ & $x_c=\infty$   \\
\hline
      & $\mathcal{E}_{0}^{\dagger}$                                                   & 123.3707084678  & 4.9511293232  
& 1.2984598320  & 0.5003910829  & 0.50000007 & 0.4999999999   \\
      & $\left(\Delta V_{0}\right)^{2}$                                               & 0.000000600468  & 0.0003747558  
& 0.0058688193  & 0.1215456043  & 0.124999   & 0.1299999999 \\ 
0     & $\left(\Delta T_{0}\right)^{2}$                                               & 0.000000600466  & 0.0003747558  
& 0.0058688193  & 0.1215456043  & 0.124999   & 0.1299999999 \\
      & $\langle T \rangle_{0}\langle V \rangle_{0}-\langle TV \rangle_{0}$     & 0.000000600466  & 0.0003747558  
& 0.0058688193  & 0.1215456043  & 0.124999   & 0.1299999999 \\
      & $\langle T \rangle_{0}\langle V \rangle_{0}-\langle VT \rangle_{0}$     & 0.000000600466  & 0.0003747558  
& 0.0058688193  & 0.1215456043  & 0.124999   & 0.1299999999 \\
\hline
      & $\mathcal{E}_{1}^{\ddag}$                                                     & 493.481633417 & 19.7745341792 
& 5.0755820152  & 1.5060815272   & 1.5000000036 & 1.499999999 \\
      & $\left(\Delta V_{1}\right)^{2}$                                               & 0.00000085445 & 0.00053374630 
& 0.0084865378  & 0.3353761814   & 0.3749997486 & 0.374999999 \\ 
1     & $\left(\Delta T_{1}\right)^{2}$                                               & 0.00000085434 & 0.00053374630 
& 0.0084865378  & 0.3353761814   & 0.3749997486 & 0.374999999 \\
      & $\langle T \rangle_{1}\langle V \rangle_{1}-\langle TV \rangle_{1}$     & 0.00000085434 & 0.00053374630 
& 0.0084865378  & 0.3353761814   & 0.3749997486 & 0.374999999 \\
      & $\langle T \rangle_{1}\langle V \rangle_{1}-\langle VT \rangle_{1}$     & 0.00000085434 & 0.00053374630 
& 0.0084865378  & 0.3353761814   & 0.3749997486 & 0.374999999 \\             
\end{tabular}
\end{ruledtabular}
\begin{tabbing} 
$^{\dagger}$Literature results \cite{roy15} of $\mathcal{E}_{0}$ for $x_{c}=0.1,0.5,1,3,5,\infty$ are: 
~123.37070846785,~4.9511293232541,~1.2984598320321,\\
~0.5003910829301,~0.5000000000768,~0.5 respectively. \\
 $^{\ddag}$Literature results \cite{roy15} of $\mathcal{E}_{1}$ for $x_{c}=0.1,0.5,1,3,5,\infty$ are: 
~493.48163341761,~19.774534179208,~5.0755820152268,\\
~0.5060815272531,~1.5000000036719,~1.5 respectively.
\end{tabbing}
\end{table}  

\section{Results and Discussion}
We shall now discuss the results under four broad category of confinement conditions \emph{viz.}, (i) impenetrable, sharp 
(ii) impenetrable, smooth (iii) penetrable, sharp (iv) penetrable, smooth, sequentially. 
\subsection{Impenetrable, sharp confinement} 
In this condition, the desired confinement effect on $v(r)$ is imposed by invoking the following form of potential: 
$v_c(r) = +\infty$ for $r > r_c$, and 0 for $r \leq r_c$,  where $r_c$ signifies radius of box. In such situation, Eq.~(16) 
needs to be solved under Dirichlet boundary condition, $\psi_{n_{r},l} (0)=\psi_{n_{r},l}(r_c)=0$. Four systems will be included, 
namely, 1DCHO, 3DCHO, CHA and SCHA, which are taken up one by one. 
\vspace{-0.2in}
\subsubsection{1DCHO}
The single-particle time-independent non-relativistic SE in 1D is ($\alpha$ is force constant):
\begin{equation}
-\frac{1}{2}\frac{d^{2}\psi_{n}}{dx^{2}}+4\alpha^{2}x^{2}\psi_{n}+v_{c}\psi_{n}= \mathcal{E}_{n}\psi_{n},
\end{equation}
where, the confining potential is defined as, $v_{c}=0$ for $x<|x_{c}|$ and $v_{c}=\infty$ for $x \ge |x_{c}|$. Here, $x_c$  
signifies confinement length. Note that we consider only the \emph{symmetric} case; while \emph{asymmetric} confinement can also 
be worked out (omitted here). Equation~(20) can be solved exactly using the boundary condition 
$\psi_{n}(-x_{c})=\psi_{n}(x_{c})=0$, to produce the following analytical closed forms for \emph{odd and even} states 
($\alpha=\sqrt{\frac{1}{8}}$, for sake of convenience),
\begin{equation}
\begin{aligned}
\psi_{e}(x)= N_{e}\ _{1}F_{1}\left[\left(\frac{1}{4}-\frac{\mathcal{E}_{n}}{4\sqrt{2}\alpha}\right),
\frac{1}{2},2\sqrt{2}\alpha x^{2}\right]e^{-\sqrt{2}\alpha x^{2}}, \\
\psi_{o}(x)=N_{o} x \ _{1}F_{1}\left[\left(\frac{3}{4}-\frac{\mathcal{E}_{n}}{4\sqrt{2}\alpha}\right),
\frac{3}{2},2\sqrt{2}\alpha x^{2}\right]e^{-\sqrt{2}\alpha x^{2}}.
\end{aligned}
\end{equation} 
In this equation, $N_{e},N_{o}$ represent normalization constant for even and odd states respectively, $\mathcal{E}_{n}$, the 
energy of respective eigenstates has been calculated accurately by an imaginary-time evolution method \cite{roy15}, while 
$_1F_1\left[a,b,x\right]$ denotes the confluent hypergeometric function. Now, the expectation values will take following forms:
\begin{equation}
\langle \hat{T} \hat{V} \rangle_{n} = \langle \hat{T} v(x) \rangle_{n} + \langle \hat{T}v_{c} \rangle_{n}= \langle \hat{T} 
v(x) \rangle_{n}.
\end{equation}  
One can make use of the property of Reimann integral to write,  
\begin{equation}
\begin{aligned}
\langle \hat{T}v_{c} \rangle_{n} & =  \int_{-\infty}^{-x_c} \psi^{*}_{n}(x)\hat{T} v_{c} \psi_{n}(x) \mathrm{d}x + 
\int_{-x_c}^{x_c} \psi^{*}_{n}(x)\hat{T} v_{c}\psi_{n}(x)
\mathrm{d}x + \int_{x_c}^{\infty} \psi^{*}_{n}(x)\hat{T} v_{c} \psi_{n}(x) \mathrm{d}x =0 \\
\end{aligned}
\end{equation}
The first and third integrals turn out as \emph{zero} because $\psi_{n}(x)=0$ when $x \ge |x_{c}|$, whereas the second integral 
becomes \emph{zero} as, $v_{c}=0$ inside the box. Similarly, 
\begin{equation}
\langle \hat{V} \hat{T} \rangle_{n} = \langle v(x)\hat{T} \rangle_{n} + \langle v_{c} \hat{T} \rangle_{n}= \langle v(x) 
\hat{T} \rangle_{n} 
\end{equation}
\begin{equation}
\langle \hat{V}^{2} \rangle_{n} = \langle (v(x)^{2} \rangle_{n} +\langle v(x) v_{c} \rangle_{n} + \langle v_{c} v(x) 
\rangle_{n} + \langle v_{c}^{2} \rangle_{n} = \langle (v(x))^{2} \rangle_{n} 
\end{equation}
\begin{equation}
\langle \hat{V} \rangle_{n} = \langle v(x) \rangle_{n} + \langle v_{c} \rangle_{n} = \langle v(x) \rangle_{n}.   
\end{equation}
Thus, for a 1DCHO, with the help of above equations, Eq.~(5) may be recast as, 
\begin{equation}
\left(\Delta \hat{T}_{n}\right)^{2} = \left(\Delta \hat{V}_{n}\right)^{2}  = \langle \hat{T} \rangle_{n} \langle v(x) \rangle_{n}
-\langle v(x)\hat{T} \rangle_{n} = \langle \hat{T} \rangle_{n} \langle v(x) \rangle_{n}
-\langle \hat{T} v(x) \rangle_{n}.
\end{equation}
Thus it is evident from Eq.~(27) that, $v_c$ has no contribution in the desired expectation values. Hence the only difference 
between the free and enclosed system is that, in the latter, the boundary has been reduced to a finite region from infinity. 
Numerical values of $\mathcal{E}_{n}$, $(\Delta \hat{T}_{n})^{2}$, $(\Delta \hat{V}_{n})^{2}$, 
$\langle T \rangle_{n}\langle V \rangle_{n}-\langle TV \rangle_{n}$ and $\langle T \rangle_{n}\langle V \rangle_{n} 
- \langle VT \rangle_{n}$ are produced in Table~I for $n=0,~1$ states of 1DCHO at six selected $x_c$ values, namely 
$0.1,~0.5,~1,~3,~5,~\infty$, that cover a large region of confinement. In all these six $x_c$, $\mathcal{E}_{0}$ and 
$\mathcal{E}_{1}$ remain in excellent agreement with available literature results as compared in \cite{roy15}, and hence not 
repeated here. However, no direct reference could be found for the expectation values to tally. It is easily noticed that, in 
both free and confined condition, Eq.~(5) is obeyed, as all the expectation values offer identical results, which validates the 
the applicability of our newly designed theorem in case of 1D CHO. Additionally, with increase in $x_c$, both 
$\Delta \hat{T}^{2}$ and $\Delta \hat{V}^{2}$ increase, which presumably occurs as the wave function delocalizes with $x_c$. 
Consequently, the difference between mean square and average values of $\hat{T},~\hat{V}$ tends to grow. 

\begingroup           
\squeezetable
\begin{table}
\caption{$\mathcal{E}_{n_r,\ell}, (\Delta V_{n_r,\ell})^{2}, (\Delta T_{n_r,\ell})^{2}, \langle T \rangle_{n_r,\ell}\langle V 
\rangle_{n_r, \ell}-\langle TV \rangle_{n_r,\ell}, \langle T \rangle_{n_r,\ell}\langle V \rangle_{n_r,\ell}-\langle VT 
\rangle_{n_r,l}$ for $1s,~1p,~2s$ states in 3DCHO at six selected $r_{c}$'s, namely $0.1, 0.5, 1, 2, 5, \infty$. 
See text for detail.}
\centering\begin{ruledtabular}
\begin{tabular}{l|l|llllll}
State & Property           &  $r_c=0.1$ & $r_c=0.5$ & $r_c=1$ & $r_c=2$ & $r_c=5$ & $r_c=\infty$  \\
\hline
      & $\mathcal{E}_{1,0}^{\P}$                                                        & 493.4816334599 & 19.774534179 
& 5.0755820153 & 1.7648164388  & 1.5000000003 & 1.499999999    \\
      & $\left(\Delta V_{1,0}\right)^{2}$                                               & 0.00000085434  & 0.0005337463 
& 0.0084865378 & 0.1211110138  & 0.3749999628 & 0.374999999  \\ 
$1s$    & $\left(\Delta T_{1,0}\right)^{2}$                                               & 0.00000085434  & 0.0005337463 
& 0.0084865378 & 0.1211110138  & 0.3749999628 & 0.374999999  \\
      & $\langle T \rangle_{1,0}\langle V \rangle_{1,0}$ $-$ $\langle TV \rangle_{1,0}$ & 0.00000085434  & 0.0005337463 
& 0.0084865378 & 0.1211110138  & 0.3749999628 & 0.374999999  \\
      & $\langle T \rangle_{1,0}\langle V \rangle_{1,0}$ $-$ $\langle VT \rangle_{1,0}$ & 0.00000085434  & 0.0005337463 
& 0.0084865378 & 0.1211110138  & 0.3749999628 & 0.374999999 \\
\hline
      & $\mathcal{E}_{1,1}^{\S}$                                                        & 1009.53830080 & 40.428276496   
& 10.282256939 & 3.246947098 & 2.5000000584 & 2.499999999  \\
      & $\left(\Delta V_{1,1}\right)^{2}$                                               & 0.0000008424  & 0.00052642239  
& 0.0084064867 & 0.129302864 & 0.6249963610 & 0.624999999 \\ 
$1p$    & $\left(\Delta T_{1,1}\right)^{2}$                                               & 0.00000084238 & 0.00052642239  
& 0.0084064867 & 0.129302864 & 0.6249963610 & 0.624999999\\
      & $\langle T \rangle_{1,1}\langle V \rangle_{1,1}-\langle TV \rangle_{1,1}$ & 0.00000084238 & 0.00052642239  
& 0.0084064867 & 0.129302864 & 0.6249963610 & 0.624999999 \\
      & $\langle T \rangle_{1,1}\langle V \rangle_{1,1}-\langle VT \rangle_{1,1}$ & 0.00000084238 & 0.00052642239  
& 0.0084064867 & 0.129302864 & 0.6249963610 & 0.624999999 \\
\hline
      & $\mathcal{E}_{2,0}^{\ddag}$                                                     & 1973.922483399 & 78.9969211469 
& 19.8996965019 & 5.5846390792 & 3.5000122149 & 3.499999999 \\
      & $\left(\Delta V_{2,0}\right)^{2}$                                               & 0.00000182     & 0.00113739969 
& 0.01815844553 & 0.2779838025 & 1.6246856738 & 1.624999999 \\ 
$2s$    & $\left(\Delta T_{2,0}\right)^{2}$                                               & 0.00000182     & 0.00113739969 
& 0.01815844553 & 0.2779838025 & 1.6246856738 & 1.624999999 \\
      & $\langle T \rangle_{2,0}\langle V \rangle_{2,0}-\langle TV \rangle_{2,0}$ & 0.00000182     & 0.00113739969 
& 0.01815844553 & 0.2779838025 & 1.6246856738 & 1.624999999 \\
      & $\langle T \rangle_{2,0}\langle V \rangle_{2,0}-\langle VT \rangle_{2,0}$ & 0.00000182     & 0.00113739969 
& 0.01815844553 & 0.2779838025 & 1.6246856738 & 1.624999999              
\end{tabular}
\end{ruledtabular}
\begin{tabbing} 
$^{\P}$Literature results \cite{roy14} of $\mathcal{E}_{1,0}$ for $r_{c}=0.1,0.5,1,3,5,\infty$ are: 
~493.48163346,~19.774534180,~5.0755820154,~1.7648164388,\\ ~1.5000000037,~1.5 respectively. \\
$^{\S}$Literature results \cite{roy14} of $\mathcal{E}_{1,1}$ for $r_{c}=0.1,0.5,1,3,5,\infty$ are: 
~1009.5383008,~40.428276496,~10.282256939,~3.2469470987,\\ ~2.5000000584,~2.5 respectively. \\
$^{\ddag}$ Literature results \cite{roy14} of $\mathcal{E}_{2,0}$ for $r_{c}=0.1,0.5,1,3,5,\infty$ are: 
1973.922483399,~78.996921147,~19.899696502,~5.5846390792,\\~3.500012215,~3.5 respectively.
\end{tabbing}
\end{table}
\endgroup

\vspace{-.25in}
\subsubsection{3DCHO}
The isotropic harmonic oscillator has the form, $v(r)=\frac{1}{2}\omega r^2$, where $\omega$ signifies the oscillation frequency. 
The \emph{exact} generalized radial wave function of a 3DCHO is mathematically expressed as \cite{montgomery07}, 
\begin{equation}
\psi_{n_{r}, \ell}(r)= N_{n_{r}, \ell} \ r^{\ell} \ _{1}F_{1}\left[\frac{1}{2}\left(\ell+\frac{3}{2}-
\frac{\mathcal{E}_{n_{r},\ell}}{\omega}\right), \left(\ell+\frac{3}{2}\right),\omega r^{2}\right] e^{-\frac{\omega}{2}r^{2}}.
\end{equation}
Here $N_{n_r, \ell}$ represents the normalization constant, $\mathcal{E}_{n_r,\ell}$ corresponds to the energy of a given state 
characterized by quantum numbers $n_r,\ell$. Note that, the levels are designated by $n_{r} +1$ and $\ell$ values, such that
$n_{r} =\ell = 0$ signifies $1s$ state. The radial quantum number $n_r$ relates to $n$ as $n = 2n_r + \ell$.

The relevant expectation values will now take following forms, 
\begin{equation}
\langle \hat{T}\hat{V} \rangle_{n_{r},\ell}= \langle \hat{T} v(r) \rangle_{n_{r},\ell} + \langle \hat{T}v_{c}(r) 
\rangle_{n_{r},\ell}= \langle \hat{T} v(r) \rangle_{n_{r},\ell}. 
\end{equation}
This occurs because $\langle \hat{T}v_{c}(r) \rangle_{n_{r},\ell}=0$, due to the wave function vanishing when $r \geq r_c$. 
A similar argument ($\langle v_{c}(r) \hat{T}\rangle_{n_{r},\ell}=0$) leads to the conclusion that, 
\begin{equation}
\begin{aligned}
\langle \hat{V}\hat{T} \rangle_{n_{r},\ell} & = & \langle v(r) \hat{T} \rangle_{n_{r},\ell} + \langle v_{c}(r) \hat{T}  
\rangle_{n_{r},\ell}= \langle  v(r) \hat{T} \rangle_{n_{r},\ell}  
\end{aligned}
\end{equation}  
Then since $\langle v(r)v_{c}(r) \rangle_{n_{r},\ell} = \langle v_{c}(r)v(r) \rangle_{n_{r},\ell} = \langle v_{c}(r)^{2} 
\rangle_{n_{r},\ell} =0$, we can write, 
\begin{equation}
\begin{aligned}
\langle \hat{V}^{2} \rangle_{n_{r}, \ell} = \langle v(r)^{2} \rangle_{n_{r},\ell} + \langle v(r)v_{c}(r) \rangle_{n_{r},\ell} + 
\langle v_{c}(r)v(r) \rangle_{n_{r}, \ell} + \langle v_{c}(r)^{2} \rangle_{n_{r},\ell} =  \langle v(r)^{2} \rangle_{n_{r},\ell}. 
\end{aligned}
\end{equation}
And finally, one can derive (since $\langle v_{c}(r) \rangle_{n_{r},\ell}=0$), 
\begin{equation}
\langle \hat{V} \rangle_{n_{r},\ell} = \langle v(r) \rangle_{n_{r},\ell} + \langle v_{c}(r) \rangle_{n_{r},\ell} =  
\langle v(r) \rangle_{n_{r},\ell}. 
\end{equation}
Thus, for a 3DCHO, Eq.~(5) remains unchanged,
\begin{eqnarray}
\langle \hat{T}^{2} \rangle_{n_{r},\ell}-\langle \hat{T} \rangle^{2}_{n_{r},\ell} & = & \langle \hat{V}^{2} 
\rangle_{n_{r},\ell}-\langle \hat{V} \rangle^{2}_{n_{r},\ell}   \\
(\Delta \hat{T}_{n_{r},\ell})^{2}  =  (\Delta \hat{V}_{n_{r},\ell})^{2} & = & \langle \hat{T} 
\rangle_{n_{r},\ell} \langle v(r) \rangle_{n_{r},\ell}
-\langle v(r)\hat{T} \rangle_{n_{r},\ell} = \langle \hat{T} \rangle_{n_{r},\ell} \langle v(r) \rangle_{n_{r},\ell}
-\langle \hat{T} v(r) \rangle_{n_{r},\ell}. \nonumber 
\end{eqnarray}
Thus we observe that, similar to 1DCHO, here also the perturbing (confining) potential makes no contribution on desired
expectation values; only the boundary in confined system gets shifted to $r_c$, from $\infty$ of the corresponding free 
counterpart. It clearly indicates the validity of Eq.~(5) in a 3DCHO. As an illustration, Table~II imprints numerically 
calculated values of appropriate expectation values, for three low-lying ($1s,~1p,~2s$) states at six selected $r_c$'s, namely, 
$0.1,~0.5,~1,~2,~5,~\infty$. This again establishes the utility of Eq.~(5) for such potential in both confined and free system, 
as evident from identical values of these quantities at all $r_c$'s--last column signifying the corresponding \emph{free} 
system. Accurate energy values are quoted from GPS results \cite{roy14}. No literature results are available for average 
values considered here. Like the 1D case, here also $(\Delta \hat{T}_{n_r,\ell})^{2},~(\Delta \hat{V}_{n_r,\ell})^{2}$ increase 
with $r_c$.   
  
\begingroup           
\squeezetable
\begin{table}
\caption{$\mathcal{E}_{n,\ell}, \left(\Delta V_{n,\ell}\right)^{2}, \left(\Delta T_{n,\ell}\right)^{2}, \langle T \rangle_{n,\ell}
\langle V \rangle_{n,\ell}-\langle TV \rangle_{n,\ell}, \langle T \rangle_{n,\ell}\langle V \rangle_{n,\ell}-\langle VT 
\rangle_{n,\ell}$ of $1s,2s,2p$ states in CHA at six selected $r_{c}$, namely, $0.1,0.2,0.5,1,5,\infty$. See text for detail.}
\centering
\begin{ruledtabular}
\begin{tabular}{l|l|llllll}
State & Property           &  $r_c=0.1$ & $r_c=0.2$ & $r_c=0.5$ & $r_c=1$ & $r_c=5$ & $r_c=\infty$  \\
\hline
      & $\mathcal{E}_{1,0}^{\P}$                                                        & 468.993038659  & 111.069858836 
& 14.7479700303  & 2.3739908660  & $-$0.4964170065 & $-$0.499999999 \\
      & $\left(\Delta V_{1,0}\right)^{2}$                                               & 308.872889980  & 80.3808359891 
& 14.5396201848  & 4.4909017616  & 1.0176222756    & 0.9999999999 \\ 
$1s$    & $\left(\Delta T_{1,0}\right)^{2}$                                               & 308.872889980  & 80.3808359891 
& 14.5396201848  & 4.4909017616  & 1.0176222756    & 0.9999999999 \\
      & $\langle T \rangle_{1,0}\langle V \rangle_{1,0}-\langle TV \rangle_{1,0}$ & 308.872889980  & 80.3808359891 
& 14.5396201848  & 4.4909017616  & 1.0176222756    & 0.9999999999 \\
      & $\langle T \rangle_{1,0}\langle V \rangle_{1,0}-\langle VT \rangle_{1,0}$ & 308.872889980  & 80.3808359891 
& 14.5396201848  & 4.4909017616  & 1.0176222756    & 0.9999999999 \\
\hline
      & $\mathcal{E}_{n,l}^{\dag}$                                                      & 1942.720354554 & 477.8516723922 
& 72.6720391904  & 16.5702560934  & 0.1412542037 & $-$0.1249999999 \\
      & $\left(\Delta V_{2,0}\right)^{2}$                                               & 925.842896028 &  236.7351455444 
& 40.5134596945  & 11.3096437104  & 0.8156705939 & 0.1874999999 \\ 
$2s$    & $\left(\Delta T_{2,0}\right)^{2}$                                               & 925.842896028 &  236.7351455444 
& 40.5134596945  & 11.3096437104  & 0.8156705939 & 0.1874999999 \\
      & $\langle T \rangle_{2,0}\langle V \rangle_{2,0}-\langle TV \rangle_{2,0}$ & 925.842896028 &  236.7351455444 
& 40.5134596945  & 11.3096437104  & 0.8156705939 & 0.1874999999 \\
      & $\langle T \rangle_{2,0}\langle V \rangle_{2,0}-\langle VT \rangle_{2,0}$ & 925.842896028 &  236.7351455444 
& 40.5134596945  & 11.3096437104  & 0.8156705939 & 0.1874999999 \\
\hline
      & $\mathcal{E}_{2,1}^{\ddag}$                                                    & 991.0075894411 & 243.10933211 
& 36.6588758801  & 8.2231383161 & 0.0075939204 & $-$0.124999999 \\
      & $\left(\Delta V_{2,1}\right)^{2}$                                              & 47.98046148 & 12.14249373     
& 2.01620344857  & 0.5370036884 & 0.0381647208 & 0.02083333333 \\ 
$2p$    & $\left(\Delta T_{2,1}\right)^{2}$                                               & 47.98046148 &  12.14249373    
& 2.01620344857  & 0.5370036884 & 0.0381647208 & 0.02083333333 \\
      & $\langle T \rangle_{2,1}\langle V \rangle_{2,1}-\langle TV \rangle_{2,1}$ & 47.98046148 &  12.14249373    
& 2.01620344857  & 0.5370036884 & 0.0381647208 & 0.02083333333 \\
      & $\langle T \rangle_{2,1}\langle V \rangle_{2,1}-\langle VT \rangle_{2,1}$ & 47.98046148 &  12.14249373    
& 2.01620344857  & 0.5370036884 & 0.0381647208 & 0.02083333333   
\end{tabular}
\end{ruledtabular}
\begin{tabbing} 
$^{\P}$Literature results \cite{roy15a} of $\mathcal{E}_{1,0}$ for $r_{c}=0.1,0.2,0.5,1,5, \infty$ are: 
~468.9930386595,~111.0698588367,~14.74797003035,~2.373990866100,\\ ~$-$0.496417006591,~$-$0.5 respectively. \\
$^{\dag}$Literature results \cite{roy15a} of $\mathcal{E}_{2,0}$ for $r_{c}=0.1,0.2,0.5,1, \infty$ are: 
~1942.720354554,~477.8516723922,~72.67203919047,~16.57025609346,\\~$-$0.125 respectively. \\
$^{\ddag}$ Literature results \cite{roy15a} of $\mathcal{E}_{2,1}$ for $r_{c}=0.1,0.2,0.5,1,\infty$ are: 
991.0075894412,~243.1093166600,~36.65887588018,~8.223138316165,\\~$-$0.125 respectively.
\end{tabbing}
\end{table}
\endgroup

\subsubsection{CHA}
We begin with the \emph{exact} wave function for CHA, which assumes the following form \cite{burrows06},   
\begin{equation}
\psi_{n, \ell}(r)= N_{n, \ell}\left(2r\sqrt{-2\mathcal{E}_{n,\ell}}\right)^{\ell} \ _{1}F_{1}
\left[\left(\ell+1-\frac{1}{\sqrt{-2\mathcal{E}_{n,\ell}}}\right),(2\ell+2),2r\sqrt{-2\mathcal{E}_{n,\ell}}\right] 
e^{-r\sqrt{-2\mathcal{E}_{n,\ell}}},
\end{equation}
with $N_{n, \ell}$ denoting normalization constant, $\mathcal{E}_{n,\ell}$ corresponding to energy of a state 
represented by $n,\ell$ quantum numbers. The pertinent expectation values can be simplified as, 
\begin{equation}
\langle \hat{T}\hat{V} \rangle_{n,\ell}= \langle \hat{T} v(r) \rangle_{n,\ell} + \langle \hat{T}v_{c}(r) \rangle_{n,\ell}= 
\langle \hat{T} v(r) \rangle_{n,\ell}. 
\end{equation}
In this instance,  $\langle \hat{T}v_{c}(r) \rangle_{n,\ell}=0$, as the wave function vanishes for $r \geq r_c$. Use of same 
argument, along with the fact that $\langle v_{c}(r) \hat{T}\rangle_{n,\ell}=0$, leads to the following,  
\begin{equation}
\begin{aligned}
\langle \hat{V}\hat{T} \rangle_{n,\ell} & = & \langle v(r) \hat{T} \rangle_{n,\ell} + \langle v_{c}(r) \hat{T}  
\rangle_{n,\ell}= \langle  v(r) \hat{T} \rangle_{n,\ell}.  
\end{aligned}
\end{equation}  
Now since $\langle v(r)v_{c}(r) \rangle_{n,\ell} = \langle v_{c}(r)v(r) \rangle_{n,\ell} = \langle v_{c}(r)^{2} 
\rangle_{n,\ell} =0$, one can write, 
\begin{equation}
\begin{aligned}
\langle \hat{V}^{2} \rangle_{n,\ell} = \langle v(r)^{2} \rangle_{n,\ell} + \langle v(r)v_{c}(r) \rangle_{n,\ell} + 
\langle v_{c}(r)v(r) \rangle_{n,\ell} + \langle v_{c}(r)^{2} \rangle_{n,\ell} =  \langle v(r)^{2} \rangle_{n,\ell}. 
\end{aligned}
\end{equation}
Again because  $\langle v_{c}(r) \rangle_{n,\ell}=0$, it follows that, 
\begin{equation}
\langle \hat{V} \rangle_{n,\ell} = \langle v(r) \rangle_{n,\ell} + \langle v_{c}(r) \rangle_{n,\ell} = \langle v(r) 
\rangle_{n,\ell}. 
\end{equation}

Thus, like the previous two systems, for CHA also, Eq.~(5) remains unchanged, i.e., 
\begin{equation}
\begin{aligned}
\langle \hat{T}^{2} \rangle_{n,\ell}-\langle \hat{T} \rangle^{2}_{n,\ell} & = \langle \hat{V}^{2} \rangle_{n,\ell}-\langle 
\hat{V} \rangle^{2}_{n,\ell} \\
\left(\Delta \hat{T}_{n,\ell}\right)^{2} = \left(\Delta \hat{V}_{n,\ell}\right)^{2} & = \langle \hat{T} \rangle_{n,\ell} 
\langle v(r) \rangle_{n,\ell}
-\langle v(r)\hat{T} \rangle_{n,\ell} = \langle \hat{T} \rangle_{n,\ell} \langle v(r) \rangle_{n,\ell}
-\langle \hat{T} v(r) \rangle_{n,\ell}.
\end{aligned}
\end{equation}
This equation implies that, CHA satisfies the results given in Eq.~(5); as before, $v_c$ has no impact on it. It has only 
introduced the boundary in a finite range. Table~III demonstrates sample values of $\mathcal{E}_{n,\ell}$, 
$(\Delta \hat{T}_{n,\ell})^{2}$, $(\Delta \hat{V}_{n,\ell})^{2}$, $\langle T \rangle_{n,\ell}\langle V 
\rangle_{n,\ell}-\langle TV \rangle_{n,\ell}$ and $\langle T \rangle_{n,\ell}\langle V \rangle_{n,\ell}-\langle VT 
\rangle_{n,\ell}$ for same low-lying ($1s,~2s,~2p$) states of previous table, in CHA at same six selected $r_c$ values, namely 
$0.1,~0.2,~0.5,~1,~5, \infty$. For sake of completeness, accurate values of $\mathcal{E}_{n,\ell}$ are reproduced from \cite{roy15a}. 
Once again, no literature results could be found to compare the numerically calculated expectation values. In both free and 
confining conditions, these results complement the conclusion of Eq.~(5). In the passing, it is interesting to note that both 
$(\Delta \hat{T}_{n,\ell})^{2}, (\Delta \hat{V}_{n,\ell})^{2}$ decrease with rise in $r_c$.

\begingroup           
\squeezetable
\begin{table}
\caption{$\mathcal{E}_{n,\ell}, \left(\Delta V_{n,\ell}\right)^{2}, \left(\Delta T_{n,\ell}\right)^{2}, \langle T 
\rangle_{n,\ell}\langle V \rangle_{n,\ell}-\langle TV \rangle_{n,\ell}, \langle T \rangle_{n,\ell}\langle V \rangle_{n,\ell}-
\langle VT \rangle_{n,\ell}$ for $1s,~2s,~2p$ states in SCHA at five sets of $\left(r_{a}, r_{b}\right)$ values. See text for 
detail.}
\centering
\begin{ruledtabular}
\begin{tabular}{l|l|lllll}
State & Property           &  $r_a=0.1,r_b=0.5$ & $r_a=0.2,r_b=1$ & $r_a=0.5,r_b=2$ & $r_a=1,r_b=5$ & $r_a=2,r_b=8$   \\
\hline
      & $\mathcal{E}_{1,0}$                                                         & 27.27172629  &  5.92023765  
& 1.34445210  &  $-$0.05806114  &  $-$0.07992493  \\
      & $\left(\Delta V_{1,0}\right)^{2}$                                           & 1.01266084   & 0.25766775   
& 0.04408223  &  0.011743288    &  0.0030965826         \\ 
$1s$    & $\left(\Delta T_{1,0}\right)^{2}$                                           & 1.01266084   & 0.25766775   
& 0.04408223  &  0.011743288    &  0.0030965826         \\
      & $\langle T \rangle_{1,0}\langle V \rangle_{1,0}-\langle TV \rangle_{1,0}$ & 1.01266084   & 0.25766777   
& 0.04408222  &  0.011743282    &  0.0030965824         \\
      & $\langle T \rangle_{1,0}\langle V \rangle_{1,0}-\langle VT \rangle_{1,0}$ & 1.01266084   & 0.25766777   
& 0.04408222  &  0.011743282    &  0.0030965824         \\
\hline
      & $\mathcal{E}_{2,0}$                                                         &  119.52182029 &  28.91900480 
&  7.87809191  & 0.85031117 & 0.325553290   \\
      & $\left(\Delta V_{2,0}\right)^{2}$                                           &  2.31747169   &  0.58308875  
&  0.097543493 & 0.02432941 & 0.00630949665  \\ 
$2s$    & $\left(\Delta T_{2,0}\right)^{2}$                                           &  2.31747169   &  0.58308875  
&  0.097543493 & 0.02432941 & 0.00630949665  \\
      & $\langle T \rangle_{2,0}\langle V \rangle_{2,0}-\langle TV \rangle_{2,0}$ &  2.31747169   &  0.58308875  
&  0.097543493 & 0.02432941 & 0.00630949665  \\
      & $\langle T \rangle_{2,0}\langle V \rangle_{2,0}-\langle VT \rangle_{2,0}$ &  2.31747169   &  0.58308875  
&  0.097543493 & 0.02432941 & 0.00630949665  \\
\hline
      & $\mathcal{E}_{2,1}$                                                         &  40.49778250 & 9.26352721 
&  2.09854297   &  0.088632364 & -0.028352228   \\
      & $\left(\Delta V_{2,1}\right)^{2}$                                           &  0.86315456  & 0.21982576 
&  0.040223458  &  0.010141187 &  0.0028634216  \\ 
$2p$    & $\left(\Delta T_{2,1}\right)^{2}$                                           &  0.86315456  & 0.21982576 
&  0.040223458  &  0.010141187 &  0.0028634216  \\
      & $\langle T \rangle_{2,1}\langle V \rangle_{2,1}-\langle TV \rangle_{2,1}$ &  0.86315456  & 0.21982576 
&  0.040223458  &  0.010141187 &  0.0028634216  \\
      & $\langle T \rangle_{2,1}\langle V \rangle_{2,1}-\langle VT \rangle_{2,1}$ &  0.86315456  & 0.21982576 
&  0.040223458  &  0.010141187 &  0.0028634216                 
\end{tabular}
\end{ruledtabular}
\end{table}
\endgroup 

\subsubsection{SCHA}
In this case, the desired confinement is accomplished by introducing the following form of potential: $v_c=\infty,$ when 
$0 < r \leq r_a$, $r \geq r_b$ and $v_c=0$ when $r_a<r<r_b$, where $r_a,~r_b$ signify the inner and outer radius respectively. 
Expectation values of such potential can then be worked out as below,  
\begin{equation}
\langle \hat{T}\hat{V} \rangle_{n,\ell}= \langle \hat{T} v(r) \rangle_{n,\ell} + \langle \hat{T}v_{c}(r) \rangle_{n,\ell}= 
\langle \hat{T} v(r) \rangle_{n,\ell},
\end{equation}
which, upon application of the property of Reimann integral provide, 
\begin{equation}
\begin{aligned}
\langle \hat{T}v_{c} \rangle_{n,\ell} & =  \int_{0}^{r_a}\psi^{*}_{n,\ell}(r) \hat{T} v_{c}\psi_{n,\ell}(r) 
r^{2}\mathrm{d}r + \int_{r_a}^{r_b} \psi^{*}_{n,\ell}(r) \hat{T} v_{c} \psi_{n,\ell}(r) r^{2}\mathrm{d}r + 
\int_{r_b}^{\infty} \psi^{*}_{n,\ell}(r)\hat{T} v_{c}\psi_{n,\ell}(r) r^{2}\mathrm{d}r \\
 & =0
\end{aligned}
\end{equation}
The first and third integrals contributes zero as wave function vanishes in these two regions. On the contrary, at $r_a<r<r_b$ 
region $v_c=0$; thus the second integral disappears. Same argument can be used to write, 
\begin{equation}
\begin{aligned}
\langle \hat{V}\hat{T} \rangle_{n,\ell} & = & \langle v(r) \hat{T} \rangle_{n,\ell} + \langle v_{c}(r) \hat{T}  \rangle_{n,\ell}= 
\langle  v(r) \hat{T} \rangle_{n,\ell}.  
\end{aligned}
\end{equation}  
The second equality hold because $\langle v_{c}(r) \hat{T}\rangle_{n,\ell}=0$. Likewise, $\langle \hat{V}^{2} \rangle_{n,\ell}$ 
may be expressed as, 
\begin{equation}
\begin{aligned}
\langle \hat{V}^{2} \rangle_{n,\ell} = \langle v(r)^{2} \rangle_{n,\ell} + \langle v(r)v_{c}(r) \rangle_{n,\ell} + 
\langle v_{c}(r)v(r) \rangle_{n,\ell} + \langle v_{c}(r)^{2} \rangle_{n,\ell} =  \langle v(r)^{2} \rangle_{n,\ell}, 
\end{aligned}
\end{equation}
since $\langle v(r)v_{c}(r) \rangle_{n,\ell}=\langle v_{c}(r)v(r) \rangle_{n,\ell}=\langle v_{c}(r)^{2} \rangle_{n,\ell}=0$. 
Next, utilizing $\langle v_{c}(r) \rangle_{n,ell}=0$, we get, 
\begin{equation}
\langle \hat{V} \rangle_{n,\ell} = \langle v(r) \rangle_{n,\ell} + \langle v_{c}(r) \rangle_{n,\ell} = \langle v(r) 
\rangle_{n,\ell}. 
\end{equation}
Collecting all these fact, we can write the final expressions for SCHA, 
\begin{equation}
\begin{aligned}
\langle \hat{T}^{2} \rangle_{n,\ell}-\langle \hat{T} \rangle^{2}_{n,\ell} & = \langle \hat{V}^{2} \rangle_{n,\ell}-\langle \hat{V} 
\rangle^{2}_{n,\ell} \\
(\Delta \hat{T}_{n,\ell})^{2} = (\Delta \hat{V}_{n,\ell})^{2} & = \langle \hat{T} \rangle_{n,\ell} 
\langle v(r) \rangle_{n,\ell}
-\langle v(r)\hat{T} \rangle_{n,\ell} = \langle \hat{T} \rangle_{n,\ell} \langle v(r) \rangle_{n,\ell}
-\langle \hat{T} v(r) \rangle_{n,\ell}.
\end{aligned}
\end{equation}

Equation~(45) explains that, similar to three previous confined cases, SCHA satisfies the results given in Eq.~(5). As before,
the role of $v_c$ is to incorporated the effect of boundary on the wave function. As mentioned earlier, closed form analytical 
solutions are unavailable in this case as yet; we have employed the GPS method to extract eigenvalues and eigenfunctions of a 
definite state. Table~IV produces the calculated values of various quantities for ground and two excited states ($1s,~2s,~2p$)
of SCHA at five chosen sets of $r_a,r_b$ values. The equality of four quantities at all shells once again justifies the 
validity of relations derived in Eq.~(5). No literature is available to compare the computed expectation values.

\begingroup           
\squeezetable
\begin{table}
\caption{$\mathcal{E}_{n,\ell}, \left(\Delta V_{n,\ell}\right)^{2}, \left(\Delta T_{n,\ell}\right)^{2}, \langle T 
\rangle_{n,\ell}\langle V \rangle_{n,\ell}-\langle TV \rangle_{n,\ell}, \langle T \rangle_{n,ell}\langle V \rangle_{n,\ell} -
\langle VT \rangle_{n,\ell}$ for $1s,2s,2p$ states in HICHA at six selected values of $r_{c}$, namely $0.1,0.2,0.5,1,5,\infty$. 
See text for detail.}
\centering
\begin{ruledtabular}
\begin{tabular}{l|l|llllll}
State & Property           &  $r_c=0.1$ & $r_c=0.2$ & $r_c=0.5$ & $r_c=1$ & $r_c=5$ & $r_c=\infty$  \\
\hline
      & $\mathcal{E}_{1,0}$                                                             & 16.80524705 & 7.43767694 
& 2.16863754 & 0.593771218$^{\P}$  & $-$0.404345971  & $-$0.499999999 \\
      & $\left(\Delta V_{1,0}\right)^{2}$                                               & 13.2294032  & 7.3539601 
& 3.6335903   & 2.30437841  & 1.1794853  & 0.9999999999 \\ 
$1s$    & $\left(\Delta T_{1,0}\right)^{2}$                                               & 13.2294032  & 7.3539601 
& 3.6335903   & 2.30437841  & 1.1794853 & 0.9999999999 \\
      & $\langle T \rangle_{1,0}\langle V \rangle_{1,0}-\langle TV \rangle_{1,0}$ & 13.2294032  & 7.3539601 
& 3.6335903   & 2.30437841  & 1.1794853 & 0.9999999999 \\
      & $\langle T \rangle_{1,0}\langle V \rangle_{1,0}-\langle VT \rangle_{1,0}$ & 13.2294032  & 7.3539601 
& 3.6335903   & 2.30437841  & 1.1794853 & 0.9999999999 \\
\hline
      & $\mathcal{E}_{2,0}$                                                             & 45.89969929 & 22.186822249 
& 8.25704419 & 3.771224646$^{\P}$  & 0.434727738 & $-$0.1249999999 \\
      & $\left(\Delta V_{2,0}\right)^{2}$                                               & 18.4752785  &  9.8452409   
& 4.4513850  & 2.5360027  & 0.78733209 & 0.1874999999 \\ 
$2s$    & $\left(\Delta T_{2,0}\right)^{2}$                                               & 18.4752785  &  9.8452409   
& 4.4513850  & 2.5360027  & 0.78733209 & 0.1874999999 \\
      & $\langle T \rangle_{2,0}\langle V \rangle_{2,0}-\langle TV \rangle_{2,0}$ & 18.4752785  &  9.8452409   
& 4.4513850  & 2.5360027  & 0.78733209 & 0.1874999999 \\
      & $\langle T \rangle_{2,0}\langle V \rangle_{2,0}-\langle VT \rangle_{2,0}$ & 18.4752785  &  9.8452409   
& 4.4513850  & 2.5360027  & 0.78733209 & 0.1874999999 \\
\hline
      & $\mathcal{E}_{2,1}$                                                             & 32.48998926 & 15.64056055 
& 5.76850468 & 2.60273839$^{\P}$ & 0.265263485 & $-$0.124999999 \\
      & $\left(\Delta V_{2,1}\right)^{2}$                                               & 1.5579056   & 0.8086356   
& 0.3486783  & 0.1899865 &  0.05526280 & 0.02083333333 \\ 
$2p$    & $\left(\Delta T_{2,1}\right)^{2}$                                               & 1.5579056   & 0.8086356   
& 0.3486783  & 0.1899865 &  0.05526280 & 0.02083333333 \\
      & $\langle T \rangle_{2,1}\langle V \rangle_{2,1}-\langle TV \rangle_{2,1}$ & 1.5579056   & 0.8086356   
& 0.3486783  & 0.1899865 &  0.05526280 & 0.02083333333 \\
      & $\langle T \rangle_{2,1}\langle V \rangle_{2,1}-\langle VT \rangle_{2,1}$ & 1.5579056   & 0.8086356   
& 0.3486783  & 0.1899865 &  0.05526280 & 0.02083333333   
\end{tabular}
\end{ruledtabular}
\begin{tabbing} 
$^{\P}$Literature results \cite{patil04} of $\mathcal{E}_{n,\ell}$ for $1s,2s,2p$ states at $r_c=1$ are: 
~0.594,~3.771,~2.603 respectively.
\end{tabbing}
\end{table}
\endgroup

\subsection{Impenetrable, smooth/homogeneous confinement} 
One such potential, $v(r)=-\frac{1}{r}+\frac{1}{2}\omega r^{2}$ was first proposed in \cite{patil04}, to mimic the quantum-dot 
structure. Later, in 2012, this was modified into a generalized form \cite{katriel12}, $v(r)=-\frac{1}{r}+
\big(\frac{r}{r_c}\big)^{k}$ $\big(k>1 $ and real; $\frac{1}{2} \omega =(\frac{1}{r_c})^{k}\big)$. At a fixed $r_c$, the 
perturbing potential takes following form,  
\[\lim_{k\to \infty}\Big(\frac{r}{r_c}\Big)^{k}=\left\{
\begin{array}{lr}
0  & \mathrm{for} \ \ \ r < r_c  \\
1  & \mathrm{for} \ \ \ r = r_c  \\
\infty & \ \mathrm{for} \ \ \ r > r_{c}.  
\end{array}
\right. 
\]
The required expectation values for this potential will then be given by, 
\begin{equation}
\begin{aligned}
\langle \hat{T}\hat{V} \rangle_{n,\ell} & = -\left\langle \hat{T} \bigg(\frac{1}{r}\bigg) \right\rangle_{n, \ell} + 
\left\langle \hat{T} \left(\frac{r}{r_c}\right)^{k}
\right\rangle_{n,\ell}, \\
\langle \hat{V}\hat{T} \rangle_{n,\ell} & = -\left\langle \left(\frac{1}{r}\right) \hat{T} \right\rangle_{n,\ell} + 
\left\langle \left(\frac{r}{r_c}\right)^{k} \hat{T} 
\right\rangle_{n,\ell},
\end{aligned}
\end{equation}
and 
\begin{equation}
\begin{aligned}
\langle \hat{V}^{2} \rangle_{n,\ell}  & =  \left\langle \frac{1}{r^{2}} \right\rangle_{n,\ell} -2 
\left\langle \frac{r^{k-1}}{r_{c}^{k}} \right\rangle_{n,\ell} +  
\left\langle \left(\frac{r}{r_c}\right)^{2k} \right\rangle_{n,\ell}, \\
\langle V \rangle_{n,\ell} & = -\left\langle \frac{1}{r} \right\rangle_{n,\ell} + \left\langle \left(\frac{r}{r_c}\right)^{k} 
\right\rangle_{n,\ell}.
\end{aligned}  
\end{equation}
Ultimately, we get the virial expression from Eq.~(5) in following form,  
\begin{equation}
\begin{aligned}
\langle \hat{T}^{2} \rangle_{n,\ell}-\langle \hat{T} \rangle^{2}_{n,\ell}  =  \left(\Delta \hat{T}_{n,\ell}\right)^{2} =
\left(\Delta \hat{V}_{n,\ell}\right)^{2}   =  \langle \hat{V}^{2} \rangle_{n,\ell}-\langle \hat{V} \rangle^{2}_{n,\ell} \\  
=\left\langle \frac{1}{r^{2}} \right\rangle_{n,\ell} -2 \left\langle \frac{r^{k-1}}{r_{c}^{k}} \right\rangle_{n,\ell} +  
\left\langle \left(\frac{r}{r_c}\right)^{2k} \right\rangle_{n,\ell} - \left\langle \frac{1}{r} \right\rangle_{n,\ell}^{2}+
2 \left\langle \frac{1}{r} \right\rangle_{n,\ell}\left\langle \left(\frac{r}{r_c}\right)^{k} \right\rangle_{n,\ell} \\ -
\left\langle \left(\frac{r}{r_c}\right)^{k} \right\rangle_{n,\ell}^{2} \\
 = \langle \hat{T} \rangle_{n,\ell} \left(\left\langle -\frac{1}{r} \right\rangle_{n,\ell} + \left\langle \left(\frac{r}{r_c}\right)^{2} 
\right\rangle_{n,\ell}\right) +\left\langle \left(\frac{1}{r}\right) \hat{T} \right\rangle_{n,\ell} - \left\langle 
\left(\frac{r}{r_c}\right)^{k} \hat{T} \right\rangle_{n,\ell} \\
 = \langle \hat{T} \rangle_{n,\ell} \left(\left\langle -\frac{1}{r} \right\rangle_{n,\ell} + \left\langle \left(\frac{r}{r_c}\right)^{2} 
\right\rangle_{n,\ell}\right) +\left\langle \hat{T} \left(\frac{1}{r}\right) \right\rangle_{n,\ell} - \left\langle \hat{T} 
\left(\frac{r}{r_c}\right)^{k}\right\rangle_{n,\ell}.
\end{aligned}  
\end{equation}

One striking difference from the previous impenetrable, sharp potentials is that, here the perturbing potential contributes in to 
the final form of expression. Now for the illustration, we choose $k=2$. In this scenario (finite positive $k$), at very small 
$r_c$, the potential blows up sharply, at $r_{c} \rightarrow \infty$ it behaves as free system, and at other definite $r_c$, 
it rises with $r$. Table~V offers sample results for $\mathcal{E}_{n,\ell}$ and related quantities of Eqs.~(5), for 
$1s,2s,2p$ states of 
HICHA at six selected $r_c$, namely $0.1,0.2,0.5,1,5,\infty$. Energies for these states, at $r_c=1$ could be compared with the 
known literature values \cite{patil04}, which shows reasonable agreement. The other computed quantities could not be compared 
due to the lack of reference values. Clearly, similar to the previous cases, these results also establish the applicability  
of our newly proposed virial-like expressions in HICHA.

\begingroup           
\squeezetable
\begin{table}
\caption{$\mathcal{E}_{n,\ell}, \ \left(\Delta V_{n,\ell}\right)^{2}, \left(\Delta T_{n,\ell}\right)^{2}, \langle T 
\rangle_{n,\ell}\langle V \rangle_{n,\ell}-\langle TV \rangle_{n,\ell}, \langle T \rangle_{n,\ell}\langle V \rangle_{n,\ell}-
\langle VT \rangle_{n,\ell}$ for $1s,2s,2p$ states in SPCHA at six selected sets of $\{V_{0}, r_{c}\}$ sets. See text for 
detail.}
\centering
\begin{ruledtabular}
\begin{tabular}{l|l|llllll}
State & Property           &  $V_{0}=0$  & $V_{0}=0 $ & $V_{0}=1 $ & $V_{0}=4 $ & $V_{0}=10 $ & $V_{0}=\infty $  \\
      &                    &  $r_c=5.77827$ & $r_c=4.87924$ & $r_c=5.72824$ & $r_c=5.75669$ & $r_c=5.49360$ & $r_c=5.80119$  \\
\hline
      & $\mathcal{E}_{1,0}^{\dag}$                                                      & $-$0.9998090 & $-$0.9990142 
& $-$0.999186 & $-$0.998703 & $-$0.997682 & $-$0.998302 \\
      & $\left(\Delta V_{1,0}\right)^{2}$                                               & 1.000433 & 1.003194       
& 1.00266 & 1.00447     & 1.00848 & 1.0047 \\ 
$1s$    & $\left(\Delta T_{1,0}\right)^{2}$                                               & 1.000433 & 1.003194       
& 1.00266 & 1.00447     & 1.00848 & 1.0047 \\
      & $\langle T \rangle_{1,0}\langle V \rangle_{1,0}-\langle TV \rangle_{1,0}$ & 1.000433 & 1.003194       
& 1.00266 & 1.00447     & 1.00848 & 1.0047 \\
      & $\langle T \rangle_{1,0}\langle V \rangle_{1,0}-\langle VT \rangle_{1,0}$ & 1.000433 & 1.003194       
& 1.00266 & 1.00447     & 1.00848 & 1.0047 \\
\hline
      & $\mathcal{E}_{2,0}$                                                             & $-$0.1578690 & $-$0.0909114   
& $-$0.035144 & 0.0128918 & 0.0818295 & 0.0434530 \\
      & $\left(\Delta V_{2,0}\right)^{2}$                                               & 0.355830  & 0.412397      
& 0.56205 & 0.64650     & 0.774448 & 0.60752  \\ 
$2s$    & $\left(\Delta T_{2,0}\right)^{2}$                                               & 0.355830  & 0.412397      
& 0.56205 & 0.64650     & 0.774448 & 0.60752 \\
      & $\langle T \rangle_{2,0}\langle V \rangle_{2,0}$ $-$ $\langle TV \rangle_{2,0}$ & 0.355830  & 0.412397      
& 0.56205 & 0.64650     & 0.774448 & 0.60752 \\
      & $\langle T \rangle_{2,0}\langle V \rangle_{2,0}$ $-$ $\langle VT \rangle_{2,0}$ & 0.355830  & 0.412397      
& 0.56205 & 0.64650     & 0.774448 & 0.60752 \\
\hline
      & $\mathcal{E}_{2,1}$                                                             & $-$0.1996605 & $-$0.1587620   
& $-$0.1406809 & $-$0.1172265 & $-$0.0832120  & $-$0.1022024 \\
      & $\left(\Delta V_{2,1}\right)^{2}$                                               & 0.032028 & 0.0413657      
& 0.046701 & 0.0637238    & 0.094861 & 0.0316629 \\ 
$2p$    & $\left(\Delta T_{2,1}\right)^{2}$                                               & 0.032028 & 0.0413657      
& 0.046701 & 0.0637238    & 0.094861 & 0.0316629 \\
      & $\langle T \rangle_{2,1}\langle V \rangle_{2,1}$ $-$ $\langle TV \rangle_{2,1}$ & 0.032028 & 0.0413657      
& 0.046701 & 0.0637238    & 0.094861 & 0.0316629 \\
      & $\langle T \rangle_{2,1}\langle V \rangle_{2,1}$ $-$ $\langle VT \rangle_{2,1}$ & 0.032028 & 0.0413657      
& 0.046701 & 0.0637238    & 0.094861 & 0.0316629   
\end{tabular}
\end{ruledtabular}
\begin{tabbing} 
$^{\dag}$Literature results \cite{koo79} of $1s$ state at these six $\{V_0, r_c\}$ pairs are: $-$0.9998, $-$0.9990, $-$0.9994,
$-$0.9990, $-$0.9980 and $-$0.9980 \\
respectively.
\end{tabbing}
\end{table}
\endgroup 

\subsection{Penetrable, sharp confinement} 
In this context, we have chosen the potential having following form,
\[v(r)=\left\{
\begin{array}{lr}
-\frac{1}{r}  & \mathrm{for} \ \ \ r < r_c  \\
V_{0}  & \mathrm{for} \ \ \ r \ge r_c,  
\end{array}
\right.
\]
where $V_{0}$ is a positive constant. It was first introduced by \cite{koo79} in 1979. The expectation values in this case, 
are given by following expressions, 

\begin{equation}
\begin{aligned}
\langle \hat{T}\hat{V} \rangle_{n, \ell} & = \big\langle \hat{T} v(r) \big\rangle_{n, \ell} = -\int_{0}^{r_c} 
\psi_{n, \ell}^{*}(r)  \hat{T} \left(\frac{1}{r}\right) 
\psi_{n, \ell} (r) \  r^{2} \mathrm{d}r + V_{0}\int_{r_c}^{\infty} \psi_{n, \ell}^{*}(r) \hat{T} \psi_{n, \ell} (r) \ 
r^{2} \mathrm{d}r \\
\langle \hat{V}\hat{T} \rangle_{n, \ell} & = \left\langle v(r) \hat{T} \right\rangle_{n, \ell} = -\int_{0}^{r_c} 
\psi_{n, \ell}^{*}(r)\left(\frac{1}{r}\right) \hat{T}\psi_{n, \ell}(r)
\  r^{2} \mathrm{d}r +V_{0}\int_{r_c}^{\infty} \psi_{n, \ell}^{*}(r) \hat{T} \psi_{n, \ell}(r) \ r^{2} \mathrm{d}r, 
\end{aligned}
\end{equation}
where the property of Reimann integral has been used. Now,
\begin{equation}
\begin{aligned}
\langle \hat{V}^{2} \rangle_{n, \ell} = \int_{0}^{r_c} \psi_{n, \ell}^{*}(r) \left(\frac{1}{r^{2}}\right) \psi_{n, \ell} (r) 
\  r^{2} \mathrm{d}r +V_{0}^{2}\int_{r_c}^{\infty} \psi_{n, \ell}^{*}(r) \psi_{n, \ell} (r) \ r^{2} \mathrm{d}r \\
-2\int_{0}^{r_c} \psi_{n, \ell}^{*}(r) \left(\frac{1}{r}\right) \psi_{n, \ell} (r) \  r^{2} \mathrm{d}r \ \ \
V_{0}\int_{r_c}^{\infty} \psi_{n, \ell}^{*}(r) \psi_{n, \ell} (r) \ r^{2} \mathrm{d}r, \\
\langle V \rangle_{n, \ell} = -\int_{0}^{r_c} \psi_{n, \ell}^{*}(r) \left(\frac{1}{r}\right) \psi_{n, \ell} (r) \  r^{2} 
\mathrm{d}r+ V_{0}\int_{r_c}^{\infty} \psi_{n, \ell}^{*}(r) \psi_{n, \ell} (r) \ r^{2} \mathrm{d}r.
\end{aligned}  
\end{equation}
After some algebra, we eventually obtain the following expressions, 
\begin{equation}
\begin{aligned}
\langle \hat{T}^{2} \rangle_{n, \ell}-\langle \hat{T} \rangle^{2}_{n, \ell}  =  \left(\Delta \hat{T}_{n, \ell}\right)^{2} =
\left(\Delta \hat{V}_{n, \ell}\right)^{2}   =  \langle \hat{V}^{2} \rangle_{n, \ell}-\langle \hat{V} \rangle^{2}_{n, \ell} \\
=\int_{0}^{r_c} \psi_{n, \ell}^{*}(r) \left(\frac{1}{r^{2}}\right) \psi_{n, \ell} (r) \  r^{2} \mathrm{d}r
+V_{0}^{2}\int_{r_c}^{\infty} \psi_{n, \ell}^{*}(r) \psi_{n, \ell} (r) \ r^{2} \mathrm{d}r \\
-2\int_{0}^{r_c} \psi_{n, \ell}^{*}(r) \left(\frac{1}{r}\right) \psi_{n, \ell} (r) \  r^{2} \mathrm{d}r \ \ \
V_{0}\int_{r_c}^{\infty} \psi_{n, \ell}^{*}(r) \psi_{n, \ell} (r) \ r^{2} \mathrm{d}r \\
-\left(-\int_{0}^{r_c} \psi_{n, \ell}^{*}(r) \left(\frac{1}{r}\right) \psi_{n, \ell} (r) \  r^{2} \mathrm{d}r+
V_{0}\int_{r_c}^{\infty} \psi_{n, \ell}^{*}(r) \psi_{n, \ell} (r) \ r^{2} \mathrm{d}r\right)^{2} \\
=\langle \hat{T} \rangle_{n, \ell} \left(-\int_{0}^{r_c} \psi_{n, \ell}^{*}(r) \left(\frac{1}{r}\right) \psi_{n, \ell} (r) \  
r^{2} \mathrm{d}r+ V_{0}\int_{r_c}^{\infty} \psi_{n, \ell}^{*}(r) \psi_{n, \ell} (r) \ r^{2} \mathrm{d}r\right) \\
+\int_{0}^{r_c} \psi_{n, \ell}^{*}(r)\left(\frac{1}{r}\right) \hat{T}\psi_{n, \ell}(r)
\  r^{2} \mathrm{d}r -V_{0}\int_{r_c}^{\infty} \psi_{n, \ell}^{*}(r) \hat{T} \psi_{n, \ell}(r) \ r^{2} \mathrm{d}r \\
=\langle \hat{T} \rangle_{n, \ell} \left(-\int_{0}^{r_c} \psi_{n, \ell}^{*}(r) \left(\frac{1}{r}\right) \psi_{n, \ell} (r) \  
r^{2} \mathrm{d}r+ V_{0}\int_{r_c}^{\infty} \psi_{n, \ell}^{*}(r) \psi_{n, \ell} (r) \ r^{2} \mathrm{d}r\right) \\
+\int_{0}^{r_c} \psi_{n, \ell}^{*}(r)  \hat{T} \left(\frac{1}{r}\right) \psi_{n, \ell} (r) \  r^{2} \mathrm{d}r - 
V_{0}\int_{r_c}^{\infty} \psi_{n, \ell}^{*}(r) \hat{T} \psi_{n, \ell} (r) \ r^{2} \mathrm{d}r
\end{aligned}
\end{equation}
Thus, analogous to HICHA, here also the perturbing term $V_{0}$ contributes to the expectation values. Table~VI presents results 
of $\mathcal{E}_{n,\ell}$, along with the respective expectation values for $1s,2s,2p$ states of SPCHA at six selected sets of 
$\{V_{0},r_c\}$ values. Very few literature results are available, except the ground-state energy, which are duly quoted; our
results display nice agreement. these results promote the validity of this virial like expression for SPCHA.     

\subsection{Penetrable, smooth/homogeneous confinement} 
One example of such potential is $v(r)=-\frac{1}{r}+v_{p,h} (r) $, where $v_{p,h} (r) =\frac{U_{0}}{e^{w\left(1-\frac{r}{r_c}\right)}+1}$, 
$U_{0}, w$ both are positive and real. Its importance and utility has been discussed in \cite{aquino13} in the context of explaining the 
interactions present in artificial atoms. The relevant expressions can be derived as follows,  
\begin{equation}
\begin{aligned}
\langle \hat{T}\hat{V} \rangle_{n, \ell} & = -\left \langle \hat{T} \left(\frac{1}{r} \right) \right \rangle_{n, \ell} + 
\left\langle \hat{T} v_{p,h} (r) \right\rangle_{n, \ell}, \\
\langle \hat{V}\hat{T} \rangle_{n, \ell} & = -\left\langle \left(\frac{1}{r}\right) \hat{T} \right\rangle_{n, \ell} + 
\left\langle v_{p,h} (r) \hat{T} \right\rangle_{n, \ell},
\end{aligned}
\end{equation}
and 
\begin{equation}
\begin{aligned}
\langle \hat{V}^{2} \rangle_{n, \ell}  & =  \left\langle \frac{1}{r^{2}} \right\rangle_{n, \ell} -2 \left\langle 
\left(\frac{1}{r}\right) v_{p,h} (r) \right\rangle_{n, \ell} +  
\left\langle v_{p,h}^2 (r) \right\rangle_{n, \ell}, \\
\langle V \rangle_{n, \ell} & = -\left\langle \frac{1}{r} \right\rangle_{n, \ell} + \left\langle 
v_{p,h} (r)\right\rangle_{n, \ell}.
\end{aligned}  
\end{equation}
Eventually we arrive at the following expression after some algebra, 
\begin{equation}
\begin{aligned}
\langle \hat{T}^{2} \rangle_{n,\ell}-\langle \hat{T} \rangle^{2}_{n,\ell}  =  (\Delta \hat{T}_{n,
\ell})^{2} =
(\Delta \hat{V}_{n,\ell})^{2}   =  \langle \hat{V}^{2} \rangle_{n,\ell}-\langle \hat{V} \rangle^{2}_{n,\ell} \\ 
=\left\langle \frac{1}{r^{2}} \right\rangle_{n,\ell} -2 \left\langle \left(\frac{1}{r}\right)
v_{p,h} (r) \right\rangle_{n,\ell} +  
\left\langle v_{p,h}^2 (r) \right\rangle_{n,\ell}   
-\left(-\left\langle \frac{1}{r} \right\rangle_{n,\ell} + \left\langle v_{p,h} (r) \right\rangle_{n,\ell}\right)^{2} \\
=\langle \hat{T} \rangle_{n,\ell}\left(-\left\langle \frac{1}{r} \right\rangle_{n,\ell} + 
\left\langle v_{p,h} (r) \right\rangle_{n,\ell}\right)
+\left\langle \left(\frac{1}{r}\right) \hat{T} \right\rangle_{n,\ell} - \left\langle v_{p,h} (r) 
\hat{T} \right\rangle_{n,\ell} \\
=\langle \hat{T} \rangle_{n,\ell}\left(-\left\langle \frac{1}{r} \right\rangle_{n,\ell} + 
\left\langle v_{p,h} (r) \right\rangle_{n,\ell}\right)
+\left\langle \hat{T} \left(\frac{1}{r}\right) \right\rangle_{n,\ell} - \left\langle \hat{T} 
v_{p,h} (r) \right\rangle_{n,\ell}.
\end{aligned}  
\end{equation}
Thus we notice that, similar to HICHA and SPCHA, here also the perturbing term $v_{p,h}(r) $ remains in the final expression.  

\begingroup           
\squeezetable
\begin{table}
\caption{$\mathcal{E}_{n, \ell}, \left(\Delta V_{n, \ell}\right)^{2}, \left(\Delta T_{n, \ell}\right)^{2}, \langle T 
\rangle_{n, \ell}\langle V \rangle_{n, \ell}$ $-$ $\langle TV \rangle_{n, \ell}, \langle T \rangle_{n, \ell}\langle V 
\rangle_{n, \ell}$ $-$ $\langle VT \rangle_{n, \ell}$ for $1s,2s,2p$ states in HPCHA at five selected $r_{c}$, namely, 
$0.1, 0.2, 0.5, 1, 5$, having $U=10,w=1000$. Last column indicates the values at $r_c=\infty$ and $U=0$. See text for detail.}
\centering
\begin{ruledtabular}
\begin{tabular}{l|l|llllll}
State & Property           &  $r_c=0.1$ & $r_c=0.2$ & $r_c=0.5$ & $r_c=1$ & $r_c=5$ & $r_c=\infty, U=0$  \\
\hline
      & $\mathcal{E}_{1,0}^{\S}$                                                             & 9.4871580 & 9.35868 & 5.25360 
& 1.1528598 & -0.4973688 & $-$0.499999999 \\
      & $\left(\Delta V_{1,0}\right)^{2}$                                               & 1.15378   & 2.4119  & 6.6390  
& 3.25938   & 1.0133575  & 0.9999999999 \\ 
$1s$    & $\left(\Delta T_{1,0}\right)^{2}$                                               & 1.15378   & 2.4119  & 6.6390  
& 3.25938   & 1.0133575  & 0.9999999999 \\
      & $\langle T \rangle_{1,0}\langle V \rangle_{1,0} -\langle TV \rangle_{1,0}$ & 1.15378   & 2.4119  & 6.6390  
& 3.25938   & 1.0133575  & 0.9999999999 \\
      & $\langle T \rangle_{1,0}\langle V \rangle_{1,0}-\langle VT \rangle_{1,0}$ & 1.15378   & 2.4119  & 6.6390  
& 3.25938   & 1.0133575  & 0.9999999999 \\
\hline
      & $\mathcal{E}_{2,0}$                                                             & 9.8734148 & 9.8593719 & 9.7728942 
& 9.029792  & 0.10745905 & $-$0.1249999999 \\
      & $\left(\Delta V_{2,0}\right)^{2}$                                               & 0.20807   & 0.346874  & 0.153805  
& 5.11808  & 0.7615043  & 0.1874999999 \\ 
2s    & $\left(\Delta T_{2,0}\right)^{2}$                                               & 0.20807   & 0.346874  & 0.153805  
& 5.11808  & 0.7615043  & 0.1874999999 \\
      & $\langle T \rangle_{2,0}\langle V \rangle_{2,0}-\langle TV \rangle_{2,0}$ & 0.20807   & 0.346874  & 0.153805  
& 5.11808  & 0.7615043  & 0.1874999999 \\
      & $\langle T \rangle_{2,0}\langle V \rangle_{2,0}-\langle VT \rangle_{2,0}$ & 0.20807   & 0.346874  & 0.153805  
& 5.11808  & 0.7615043  & 0.1874999999 \\
\hline
      & $\mathcal{E}_{2,1}$                                                             & 9.8749992211 & 9.87497532482 
& 9.869939026 & 4.980371 & $-$0.011992 & $-$0.124999999 \\
      & $\left(\Delta V_{2,1}\right)^{2}$                                               & 0.02083685   & 0.0209243374  
& 0.04006099  & 0.36608  & 0.03609 & 0.02083333333 \\ 
2p    & $\left(\Delta T_{2,1}\right)^{2}$                                               & 0.02083685   & 0.0209243374  
& 0.04006099  & 0.36608  & 0.03609 & 0.02083333333 \\
      & $\langle T \rangle_{2,1}\langle V \rangle_{2,1}-\langle TV \rangle_{2,1}$ & 0.02083685   & 0.0209243374  
& 0.04006099  & 0.36608  & 0.03609 & 0.02083333333 \\
      & $\langle T \rangle_{2,1}\langle V \rangle_{2,1}-\langle VT \rangle_{2,1}$ & 0.02083685   & 0.0209243374  
& 0.04006099  & 0.36608  & 0.03609 & 0.02083333333   
\end{tabular}
\end{ruledtabular}
\begin{tabbing} 
$^{\S}$Literature results \cite{aquino13} of $\mathcal{E}_{1,0}$ for $r_c=0.1,~0.2,~0.5,~1.0,~5.0,~\infty$ are: 
9.4973,~9.3620,~5.2456,~1.1761,~$-$0.4947,~$-$0.5000  \\
respectively.
\end{tabbing}
\end{table}
\endgroup 

In order to explain the result for HPCHA, we have taken $w=1000$ and $U_{0}=10$ as the potential parameters. Table~VII reports the 
calculation of $\mathcal{E}_{n, \ell}$, $\left(\Delta \hat{T}_{n, \ell}\right)^{2}$, $\left(\Delta \hat{V}_{n, \ell}\right)^{2}$, 
$\langle T \rangle_{n, \ell}\langle  V \rangle_{n, \ell}-\langle TV \rangle_{n, \ell}$ and $\langle T \rangle_{n, \ell}\langle V 
\rangle_{n, \ell}-\langle VT \rangle_{n, \ell}$ for $1s,2s,2p$ states at five selected $r_c$, namely $0.1,0.2,0.5,1,5$. Apart
form that, the last column clearly suggests at $r_c \rightarrow \infty$ and $U \rightarrow 0$ this system merges to FHA. These results, 
like the previous cases, demonstrate that relation (5) is valid for HPCHA. Ground-state energies at all these $r_c$'s are compared with the 
available literature results. No further comparison could be made due to lack of data.   

\section{Future and Outlook}
{\color{red} A new virial-like relation ($(\Delta \hat{T}_{n})^{2}=(\Delta \hat{V}_{n})^{2} $) has been proposed for \emph{free}, and \emph{confined} quantum 
systems, by invoking SE and HVT. This can be used as an essential condition for an eigenstate to obey. Besides this, Eq.~(5) in its complete form has 
been proven to be a sufficient condition for these bound, stationary states to obey.} Generalized expressions have been derived 
for impenetrable, penetrable, and shell-confined quantum systems along with the sharp and smooth situations. The change in boundary condition 
does not influence the form of these relations. Their applicability has been tested and verified by doing pilot calculations on quantum 
harmonic oscillator and H atom--a total of seven different confining potentials, as well as the respective free systems. In all cases these 
conditions are found to be obeyed. {\color{red} In impenetrable and sharp (hard) confinement condition the perturbing term is not contributing in the
final expression. But in impenetrable-smooth, penetrable-sharp, and penetrable-smooth cases it participates in the eventual form.} There are several open 
questions that may lead to important conclusions, and require further scrutiny, such as, use of these sufficient conditions in the context of determining 
optimized wave function for various quantum systems, in both ground and excited states. {\color{red}Importantly, one can perform unconstrained optimization 
(without employing the orthogonality criteria) of trial states by adopting this condition.} A parallel inspection on many-electron systems would be highly desirable.      
      
\section{Acknowledgement}
Financial support from DST SERB, New Delhi, India (sanction order: EMR/2014/000838) is gratefully acknowledged. 	

\end{document}